\documentstyle[12pt]{article}
\normalbaselines 
\newcommand{\beq}[1]{\begin{equation}\label{#1}}
\newcommand{\eeq}{\end{equation}}
\newcommand{\bear}[1]{\begin{eqnarray}\label{#1}}
\newcommand{\ear}{\end{eqnarray}}

\newcommand{\np}{ {\newpage } }

\newcommand{\Iff}{ {\Leftrightarrow } }

\newcommand{\imp}{\ {\Rightarrow }\ }

\newcommand{\R}{ \mbox{\rm I$\!$R} }
\newcommand{\Diff}{ \mbox{\rm Diff} }
\newcommand{\Out}{ \mbox{\rm Out} }
\newcommand{\Inn}{ \mbox{\rm Inn} }
\newcommand{\Aut}{ \mbox{\rm Aut} }
\newcommand{\obs}{ \mbox{\rm obs} }

\newcommand{\tr}{ \mbox{\rm tr} }

\begin{document} 
\vspace*{-1.0cm}
\centerline{\mbox{\hspace*{15cm}Preprint IPM-96}}
\vspace*{1.3cm}
\begin{center} 
{\large
 {\bf A regularizing commutant duality for 
\vspace*{ 0.5cm}  \\
a kinematically covariant partial ordered net of observables
\footnote{\bf\em This work was financially supported by a 
DAAD fellowship (M.R.). \vspace*{0.13cm}  }
\\}}  

\vskip 1.1cm 

{\large {\bf 
Martin Rainer\footnote { e-mail:
mrainer@aip.de, mrainer@physics.ipm.ac.ir}
\S\dag}} 
and 
{\large {\bf 
Hadi Salehi
\footnote { e-mail:
salehi@netware2.ipm.ac.ir}
\S\ddag}}\\
\vskip 1.1 cm  

{\S Institute for Studies in Physics and Mathematics} \\
P.O.Box 19395-5531, Tehran, Iran\\ 
\vskip 0.33 cm 

{\dag Institut f\"ur Mathematik} \\
{ Universit\"at Potsdam, PF 601553} \\
{D-14415 Potsdam, Germany}\\
\vskip 0.33 cm 

\ddag Arnold Sommerfeld Institute for Mathematical Physics\\ 
TU Clausthal, Leibnizstr. 10 \\ 
D-38678 Clausthal-Zellerfeld, Germany 

\end{center}

\vspace{1cm}
\begin{abstract}
We consider a net of $*$-algebras,  locally around any point of observation,
equipped with a natural partial order related to the isotony property. 
Assuming the underlying manifold of
the net to be a differentiable, this net shall be kinematically covariant 
under general diffeomorphisms. However, the dynamical relations,
induced by the physical state defining the related net of (von Neumann)
observables, are in general not covariant under all diffeomorphisms, 
but only under the subgroup of dynamical symmetries. 
\\
We introduce algebraically both, IR and UV cutoffs, and assume that these
are related by a commutant duality. The latter, having strong 
implications on the net, allows us to identify a $1$-parameter
group of the dynamical symmetries with the group of outer modular
automorphisms. 
\\
For thermal equilibrium states, the modular dilation parameter
may be used locally to define the notions of both, 
time and a causal structure.
\end{abstract}
\vspace{1cm}
\np
About 30 years ago Ekstein \cite{Ek} introduced the 
concept of {\em presymmetry},
as the remaining kinematical effect of covariance (there called
space-time symmetry) for the {\em observation proceedures}
even when it is broken for the {\em observables}.

The {observation proceedures} represent the abstract kinematical framework
for {possible} preparations of measurements, while the observables
encode the kinds of questions we can ask from the physical system.
It is intuitively clear, that the same question can be asked in many 
different forms, i.e. in general there are many possible preparations
of measurement for the same observable.

The covariance group of the observation proceedures reflects their 
general structure. The more sophisticated the structure of the
observation proceedures, the smaller the covariance group will be in general.
E.g. in \cite{MR} the kinematical observation proceedures are given
by a network of discrete vertices of a specific Riemannian surface
embedded in a $3+1$-dimensional space-time $M$, whence the covariance
group is only that subgroup of $\Diff(M)$ which leaves this structure
invariant. In general it is a difficult question, how much structure
might be put on the observation proceedures. 

In a concrete observation the kinematical covariance will be broken.
So in \cite{MR} a concrete local observation requires the 
explicit selection of one of many apriori equivalent vertices,
whence it breaks the covariance which holds for the network of vertices
as a whole.
In the examples of \cite{Ek} the kinematical covariance was assumed to be 
broken in a concrete observation by a dynamical interaction with external 
fields. 
 
We may say that a {presymmetry} exists if, irrespectively of the loss of 
covariance in a concrete observation, the action
of the covariance group is still welldefined on the observation proceedures.
In any case, the loss of covariance in a concrete observation is related
to a specific structure of the state of the physical system.
Hence, in the following we consider the loss of general covariance
as directly induced by the physical state itself.

Let us examine now the consequences of this breaking of general covariance 
within an algebraic approach  to generally covariant quantum field theory,
which has been proposed in \cite{FrHa} and further considered
in \cite{Sal,Sal2}.
{}From the principle of locality, which
is at the heart of the standard algebraic 
approach to quantum field theory \cite{HaKa},
we keep the assertions that the observation procedures correspond to 
{possible} preparations of localized measurements in finite regions. 
However, here we do not want to specify
a notion of time or a causal structure a priori. 
It was shown in \cite{Ba1} for
a net of subalgebras of a Weyl algebra that, it is indeed possible
to work with a flexible notion of causality rather than a rigidly given one. 
In the same spirit, here we do not impose any a priori 
causal relations between observables on different regions. 

In principle, it is even possible \cite{Ba2}
to construct a net of algebras, together with its 
underlying Hausdorff topological space $M$, by the partial 
order via inclusion of (the set of subsets of) the algebras
themselves. Although we find this approach,
where the net and its underlying manifold $M$ are derived just from
the algebras, very appealing, it is beyond the scope of the present 
investigations.
For our examination on the dynamical symmetries we need 
a differentiable structure on $M$. It might be, that even this
structure can be derived in not too ambigious manner
(cf. some recent discussion in \cite{Brans}) with the help of some
algebraic methods related to noncommutative geometry \cite{Co},
help  one. However,
in the present approach we just work a priori
with a net of $*$-algebras on an underlying differentiable 
manifold $M$.

Such a net net associatiates to each open set 
${\cal O} \in M$ a $*$-algebra ${\cal A}({\cal O})$
such that isotony, 
\beq{iso}
{\cal O}_1\subset {\cal O}_2\imp
{\cal A}({\cal O}_1)\subset {\cal A}({\cal O}_2),
\eeq
holds ($\subset$ here always denotes a proper, nonidentical inclusion).
Selfadjoint elements of ${\cal A}({\cal O})$ 
may be interpreted as {\em observation procedures},
i.e. possible prescriptions for laboratory measurements in $\cal O$. 

There should not be any a priori relations between observation procedures 
associated with disjoint regions. In other words, the net
${\cal A}:={\bigcup_{\cal O}} {\cal A}({\cal O})$ has to be
free from any relations which exceed its mere definition.\\

This interpretation allows us to extend the $\Diff(M)$ covariance  
{}from the underlying manifold $M$ to the net of algebras,
on which $\Diff(M)$ then acts by automorphisms, 
i.e. each diffeomorphism
$\chi\in \Diff(M)$ induces an automorphism 
$\alpha_{\chi}$ of the observation proceedures such that
\begin{equation}
\alpha_{\chi}({\cal A}({\cal O}))={\cal A}(\chi({\cal O})).
\label{I1}  
\end{equation}
The state of a physical system is mathematical described by a 
positive linear functional  ${\omega}$ on $\cal A$. 
Given the state ${\omega}$, one
gets via the GNS construction a representation $\pi^{\omega}$ of $\cal A$ 
by a net of operator algebras in a Hilbert space ${\cal H}^{\omega}$ with 
a cyclic vector $\Omega^{\omega}\in {\cal H}^{\omega}$. The
GNS representation $(\pi^{\omega}, {\cal H}^{\omega}, \Omega^{\omega})$ 
of any state $\omega$ has a socalled folium ${\cal F}^{\omega}$, 
given as the family of those states $\omega_\rho:=\tr\rho\pi^{\omega}$ 
which are defined by positive trace class
operators $\rho$ on ${\cal H}^{\omega}$. 

Once a physical state $\omega$ has been specified, one can consider in each
algebra ${\cal A}({\cal O})$ the equivalence relation
\begin{equation}
A\sim B \ \ :\Iff \ \ 
{\omega}^{\prime}(A-B)=0,\ \ \forall  {\omega}^{\prime}\in{\cal F}^{\omega}.
\end{equation}
These equivalence relations generate a two-sided ideal 
${\cal I}^{\omega}({\cal O}):=\{a\in{\cal A}({\cal O})\vert
{\omega}^{\prime}(A)=0\}$ in ${\cal A}({\cal O})$.
The algebra of observables 
${\cal A}^{\omega}_{\obs}({\cal O}):=\pi^{\omega}({\cal A}({\cal O}))$ may
be constructed from
the algebra of observation procedures ${\cal A}({\cal O})$ by taking 
the quotient
\begin{equation}
{\cal A}^{\omega}_{\obs}({\cal O}):=
{\cal A}({\cal O})/{\cal I}^{\omega}({\cal O}).
\end{equation}
Since any diffeomorphism $\chi\in \Diff(M)$ induces an 
automorphism $\alpha_{\chi}$ of the observation proceedures,
one may ask whether, for a given state $\omega$,  
the action of $\alpha_{\chi}$ will leave the net 
${\cal A}^{\omega}_{\obs}:=
{\bigcup_{\cal O}} {\cal A}^{\omega}_{\obs}({\cal O})$
of observables invariant, with an action of the form
\begin{equation}
\alpha_{\chi}({\cal R}^{\omega}_{\obs}({\cal O}))=
{\cal R}^{\omega}_{\obs}(\chi({\cal O})).
\label{D1}
\end{equation}
In order for this to be possible, the ideal
${\cal I}^{\omega}({\cal O})$ must be transformed covariantly, i.e.
the diffeomorphism ${\chi}$ must satisfy
\begin{equation}
\alpha_{\chi}({\cal I}^{\omega}({\cal O}))=
{\cal I}^{\omega}(\chi({\cal O})).
\label{D2}
\end{equation}
Hence, the algebra of 
observables, constructed with respect to
the folium ${\cal F}^\omega$, does no longer exhibit the 
kinematical  $\Diff(M)$ symmetry of the observation proceedures. 
The symmetry of the observables is dependent on (folium of) the state
$\omega$. 
Therefore, the selection of a folium of states
${\cal F}^\omega$, induced by the actual choice of a state $\omega$,
results immediately in a breaking of the $\Diff(M)$ symmetry.
The resulting effective symmetry group, also briefly called  
the {\em dynamical group} of the state $\omega$,
is given by the subgroup of those diffeomorphisms which satisfy 
the constraint condition (\ref{D2}).  
An automorphisms $\alpha_{\chi}$ is called {\em dynamical}
(w.r.t. the given state $\omega$) if it satisfies (\ref{D2}).

The remaining dynamical symmetry group, depending
on the folium ${\cal F}^\omega$ of states related to $\omega$,
has two main aspects which we have to examine
if we actually want to specify the physically admissible states:
Firstly, it is necessary to specify its state dependent automorphic 
algebraic action on the net of observables. Secondly, we have to
find a geometric interpretation for the group and its action on $M$.

If we consider the dynamical group as an {\em inertial}, 
and therefore global, manifestation of dynamically ascertainable 
properties of observables, 
then its (local) action should be correlated with (global) 
operations on the whole net of observables.
This implies that at least some of the dynamical
automorphisms $\alpha_\chi$ are not inner.
(For the case of causal nets of algebras 
it was actually already shown in \cite{Wo} 
that, under some additional assumptions,
the automorphisms of the algebras are in general not inner.)

Note that we might consider instead of the net of observables
${\cal A}^{\omega}_{\obs}$  the net of associated von Neumann algebras
${\cal R}^{\omega}_{\obs}({\cal O})$, which can be defined even for
unbounded ${\cal A}^{\omega}_{\obs}$, if we take 
{}from the modulus of the von Neumann closure  
${{\cal A}^{\omega}_{\obs}}''$ 
all its spectral projections \cite{FrHa}.
Then the isotony (\ref{iso}) induces a likewise isotony of
on the net 
${\cal R}^{\omega}_{\obs}:=
{\bigcup_{\cal O}} {\cal R}^{\omega}_{\obs}({\cal O})$
of von Neumann observables.

In the following we want to exhibit a possibility to introduce
in an algebraic manner both IR and UV cutoff regularizations simultaneously,
together with a partial ordering on the net of von Neumann observables. 
Let us consider nonzero open sets 
${\cal O}^{x}_s$, located around an arbitrary point $x\in M$,
and continuously parametrized by a real parameter $s$ with $0<s<\infty$ 
such that
\begin{equation}
s_1<s_2\imp{\cal O}^{x}_{s_1}{\subset} {\cal O}^{x}_{s_2}
\label{inc}
\end{equation}
and 
\begin{equation}
s\to 0\imp{\cal O}^{x}_{s}{\to} \emptyset.
\end{equation}
On open sets with  parameter $s$ restricted to 
$0<s_{\min}<s<s_{\max}<\infty$, 
the isotony property 
implies that 
\begin{equation}
{\cal R}^{\omega}_{\obs}({\cal O}^{x}_{s_{\min}})
{\ \subset\ } {\cal R}^{\omega}_{\obs}({\cal O}^{x}_{s})
{\ \subset\ } {\cal R}^{\omega}_{\obs}({\cal O}^{x}_{s_{\max}}).
\end{equation}
The key step is now to impose a commutant duality relation between
the inductive limits given by the minimal and maximal algebras,
\begin{equation}
{\cal R}^{\omega}_{\obs}({\cal O}^{x}_{s_{\min}})
={{\cal R}^{\omega}_{\obs}({\cal O}^{x}_{s_{\max}})}',
\label{dmin}
\end{equation}
where ${\cal R}'$ denotes the commutant of ${\cal R}$. 
Then the bicommutant theorem (${\cal R}''={\cal R}$) 
implies that likewise also
\begin{equation}
{\cal R}^{\omega}_{\obs}({\cal O}^{x}_{s_{\max}})
={{\cal R}^{\omega}_{\obs}({\cal O}^{x}_{s_{\min}})}'.
\label{dmax}
\end{equation}
If we now demand that all maximal (or all minimal) algebras
are isomorphic to each other, independently of the choice of $x$ and the
open set ${\cal O}^{x}_{s_{\max}}$ 
(resp. ${\cal O}^{x}_{s_{\min}}$),
then by (\ref{dmin}) (resp. (\ref{dmax})) also all minimal (resp. maximal)
algebras are isomorphic to each other.  
We then denote the universal minimal resp. maximal algebra as 
${\cal R}^{\omega}_{{\min}}$ and ${\cal R}^{\omega}_{{\max}}$ 
respectively.
Note that the duality (\ref{dmin}) implies that 
${\cal R}^{\omega}_{{\min}}$ is Abelian, and
${\cal R}^{\omega}_{{\max}}$ has necessarily a nontrivial center 
within ${\cal R}^{\omega}_{{\obs}}$.  

By isotony and (\ref{inc}), the mere existence of 
${\cal R}^{\omega}_{\min}$ 
resp. ${\cal R}^{\omega}_{\max}$
fixes  already a common size (as measured by the parameter $s$) of
all sets ${\cal O}^{x}_{s_{\min}}$ resp. ${\cal O}^{x}_{s_{\max}}$
independently of $x\in M$.    
So in this case $s_{\min}$ and $s_{\max}$ really denote 
an universal short resp.large scale cutoff.

The number $s\in]s_{\min},s_{\max}[$
parametrizes the partial order of the net of algebras
spanned between the inductive limits 
${\cal R}^{\omega}_{\min}$ and ${\cal R}^{\omega}_{\max}$.
In our theory, where the lower end of the net is Abelian, 
observations on minimal regions are expected to be rather classical,
while, for increasing size, quantum (field) theory might be
rather nontrivial.
In \cite{Wo} it was shown that for causal nets
the algebras of QFT are not Abelian and not finite-dimensional.
The Abelian character of  algebras at the lower end might 
find a natural explanation in a classical, rather than a full QFT behaviour,
at short distances. For gravity it has been indeed proposed that,
at (ultra-)short distances, it might be described in terms of an underlying 
classical kinetical theory.

If we consider the algebraic UV and IR cutoffs as introduced above,
it should be clear that only those regions (\ref{dc})
of size $s\in [s_{\min},s_{\max}]$ are admissable for measurement.  
The commutant duality between 
${\cal R}^{\omega}_{\min}$ and ${\cal R}^{\omega}_{\max}$
inevitably yields large scale correlations 
in the structure of any physical state $\omega$ on
any admissable region ${\cal O}^{x}_{s}$ of measurement at $x$.
Indeed, by isotony, the
annihilation 
of the GNS vector $\Omega^{\omega}$ in 
${\cal R}^{\omega}_{\max}$ 
implies automatically its likewise annihilation on 
${\cal R}^{\omega}_{\min}={{\cal R}^{\omega}_{\max}}^{\prime}$.
Hence,
if  $\Omega^{\omega}$ is cylic for
${\cal R}^{\omega}_{\max}$, and hence 
separating for ${{\cal R}^{\omega}_{\max}}'$, 
it should also be separating for 
${\cal R}^{\omega}_{\max}$.
So $\Omega^{\omega}$ is a cyclic and separating vector for 
${\cal R}^{\omega}_{max}$, and by isotony also for any
local von Neumann algebra ${\cal R}^{\omega}_{\obs}({\cal O}^{x}_{s})$.

As a further consequence, on any region ${\cal O}^{x}_{s}$, 
the Tomita operator $S$ and and its conjugate $F$ 
can be defined densely by
\begin{equation}
S A \Omega^{\omega}:= A^{*} \Omega^{\omega} \ \ \mbox{for}\ \ A\in 
{\cal R}^{\omega}_{\obs}({\cal O}^{x}_{s})
\label{T0}
\end{equation}
\begin{equation}
F B \Omega^{\omega}:=B^{*} \Omega^{\omega}
 \ \ \mbox{for}\ \ B\in 
{{\cal R}^{\omega}_{\obs}({\cal O}^{x}_{s})}' . 
\label{T1}
\end{equation}
The  closed Tomita operator $S$ has a polar decomposition 
\begin{equation}
S=J\Delta^{1/2},
\label{T2}\end{equation}
where  $J$ is antiunitary and $\Delta:=FS$ is the self-adjoint, positive 
modular operator.
The Tomita-Takesaki theorem \cite{Ha} provides us with a one-parameter 
group of state dependent automorphisms $\alpha^{\omega}_t$ on 
${\cal R}^{\omega}_{\obs}({\cal O}^{x}_{s})$,
defined by
\begin{equation}
\alpha^{\omega}_t (A)= \Delta^{-it}\ A\ \Delta^{it},   \ \ \mbox{for}\ \ 
A\in{\cal R}^{\omega}_{\max}.
\label{T3}
\end{equation}
So, as a consequence of commutant duality and isotony assumed above,  
we obtain here a strongly continous unitary 
implementation  of the modular group of $\omega$,
which is defined by  the $1$-parameter family of automorphisms (\ref{T3}),
given as conjugate action of operators
$e^{-it\ln\Delta}$, ${t\in\R}$.
By (\ref{T3}) the modular group, for a state $\omega$  
on the net of von Neumann algebras, 
defined by ${\cal R}^{\omega}_{\max}$,  
might be considered it as a $1$-parameter subgroup of the dynamical group.
Note that, with Eq. (\ref{T1}), in general,
the modular operator $\Delta$ is not located on
${\cal O}^{x}_{s}$. Therefore,  in general, the modular automorphisms 
(\ref{T3}) are not inner.
It is known (see e.g. \cite{BaWo}) that the modular automorphisms act as 
{inner} automorphisms, iff the von Neumann algebra 
${\cal R}^{\omega}_{\obs}({\cal O}^{x}_{s})$
generated by $\omega$
contains only semifinite factors, 
i.e. factors of type I and II. In this case  $\omega$ is a semifinite
trace.  

Above we considered concrete von Neumann algebras
${\cal R}^{\omega}_{\obs}({\cal O}^{x}_{s})$, which are in fact 
operator representations of an abstract von Neumann algebra
${\cal R}$ on a GNS Hilbertspace ${\cal H}^{\omega}$ w.r.t.
a faithful normal state $\omega$.
In general, different faithful normal states generate different 
concrete von Neumann algebras and different modular automorphism groups
of the same abstract  von Neumann algebra.
Let $\omega_1$ and $\omega_2$ be two different faithful normal states 
on a von Neumann algebra ${\cal R}$, and   
$\Omega_1$ resp. $\Omega_2$ the corresponding cyclic and separating
vectors in the corresponding GNS representation $\pi_1$ resp. $\pi_2$. 
Then the  unitary cocycle theorem \cite{BaWo} 
asserts that there exists
a strongly continuous $1$-parameter family of unitary operators
$U(t)$, which satisfy the cocycle condition 
\begin{equation}
U(t+s)=U(t)\alpha^{\omega_1}_t(U(s))   \ \ \mbox{for}\ \ t,s\in\R,
\label{UCC}
\end{equation}
and relate the modular group of $\omega_2$ to that of $\omega_1$, 
\begin{equation}
\alpha^{\omega_2}_t(A)=U(t) \alpha^{\omega_1}_t(A) U^{*}(t)
\ \ \mbox{for}\ \ A\in{\cal R}, \ t\in\R.
\label{UCC2}
\end{equation}
Any two modular groups related by (\ref{UCC2}) are called outer equivalent.
The unitarities $U(t)$ are are in fact given as 
$U(t):=e^{-it H_{\Omega_1,\Omega_2}}$
with a relative Hamiltonian
\beq{rH}
H_{\Omega_1,\Omega_2}:=\ln(\Delta_{\Omega_1,\cdot}
/\Delta_{\Omega_2,\cdot}),
\eeq
where  $\Delta_{\Omega_1,\Omega_2}$ is the relative modular operator
of the relative Tomita operator $S_{\Omega_1,\Omega_2}$, densely defined by
\begin{equation}
S_{\Omega_1,\Omega_2} \pi_2(A) \Omega_2:= \pi_1(A^{*}) \Omega_1
\ \ \mbox{for}\ \ A\in {\cal R}.
\label{rT0}
\end{equation}
With (\ref{UCC2}), the operators $U(t)$ are the intertwiners 
between the two modular groups. They yield the called Radon-Nikodym cocycles
$(D\omega_1:D\omega_2)$,
whence they are also denoted as $(D\omega_1:D\omega_2)(t):=U(t)$. 
The cocycles  satisfy the chain rule 
$(D\omega_1:D\omega_2)(D\omega_2:D\omega_3)=(D\omega_1:D\omega_3)$.

If $\alpha^{\omega_1}_t$ is inner, it can be implemented by
unitarities $e^{-itH_{\Omega_1}}$. Then (\ref{UCC2}) implies that 
there exists also a Hamiltonian $H_{\Omega_2}$, such that 
the relative Hamiltonian (\ref{rH}) takes
the form $H_{\Omega_1,\Omega_2}=H_{\Omega_1}-H_{\Omega_2}$,
whence $(D\omega_1:D\omega_2)$ is a coboundary in the group cohomology. 

The outer modular automorphisms form the cohomology group  
$\Out {\cal R}:=\Aut {\cal R}/\Inn {\cal R}$
of modular automorphisms modulo inner modular automorphisms, 
characteristic
for the types of factors contained in von Neumann algebra $\cal R$.
Per definition $\Out {\cal R}$ is trivial for inner automorphisms.
Factors of type III${}_1$ yield $\Out {\cal R}=\R$.

Hence, in the case of thermal equilibrium states, 
corresponding to factors 
of type III${}_1$ (see \cite{Ha}), there is a distinguished
$1$-parameter group of outer modular automorphisms,
which is a subgroup of the dynamical group. 

Locking for a geometric interpretation for this subgroup, 
parametrized by $\R$, it should not be a coincidence
that our partial order defined above could be parametrized
by an open interval $]s_{\min},s_{\max}[$, which is in fact  diffeomorphic
to $\R$. Therefore the dilations of the open sets
should correspond to the $1$-parameter group of outer modular automorphisms
of thermal equilibrium states.    
The effect of large scale correlations thus becomes
related to a thermal behavior of our localized states.
A local equilibrium state might be characterized
as a KMS state  (see \cite{BrRo,Ha}) over the algebra of observables 
on a double cone, whence (in agreement with
the suggestions of \cite{CoRo}) the $1$-parameter modular group in the KMS
condition might be related to the time evolution.
Note that, for double cones, a partial order
may be induced from a split property of the algebras.   

A geometric action of the modular group
might be obtained by relating the thermal time
to the geometric notion of dilations of the open sets. 
For any $x\in M$, the parameter $s$ measures the extension
of the sets ${\cal O}^{x}_{s}$. 
As accessability regions for a local 
measurement in $M$, these sets naturally increase with time. 
Hence it is natural to suggest that the parameter $s$ might be
related to the thermal time $t$.
used to introduce a notion of time $t<s$ within a set 
${\cal O}^{x}_{s}$.

For the ultralocal case (without UV cutoff), in \cite{Keyl} 
a construction of the causal structure for a space-time was 
bases on the corresponding net of operator algebras.

Nevertheless, let us for the moment still consider an apriori given 
underlying manifold $M$ of the net.
Locally around any point $x\in M$ we may induce  open 
double cones  as the 
pullback of the standard double cone, which in fact is 
the conformal model of  Minkowski space.
These open double cones then carry natural
notions of time and causality, 
which are preserved under dilations.  
Therefore it seems natural to introduce locally around any $x\in M$ 
a causal structure and time by specializing the open sets to be
open double cones ${\cal K}^{x}_{s}$ located at $x$, with
timelike extension $2s$ between the ultimate past event $p$ and
the ultimate future event $q$ involved in any measurement in 
${\cal K}^{x}_{s}$ at $x$ 
(time $s$ between $p$ and $x$, and likewise between $x$ and $q$). 
Since the open double cones form a basis for the local topology of $M$, 
we might indeed consider equivalently the net of algebras located
on open sets 
\beq{dc}
{\cal O}^{x}_{s}:={\cal K}^{x}_{s}.
\eeq
It is a difficult question, under which consistency conditions  
a local notion of time and causality might be extended, 
{}from nonzero environments of individual points to global regions.
This will not be discussed here, but elsewhere \cite{RaSa}.

However, if we assume the presence of factors of type
III${}_1$ in our von Neumann algebras, or likewise the
existence of local equilibrium states, the choices
for time and causality, made above on the basis of
a partial order given by dilations which could algebraically
be related to a commutant duality, are apparently natural. 


\end{document}